\RequirePackage{ifpdf}
\documentclass[11pt,a4paper]{article}
\pdfoutput = 1
\usepackage{jheppub}
\usepackage{mathrsfs}
\usepackage{dcolumn}
\usepackage{bm}
\usepackage{color}
\usepackage{ulem}

\providecommand{\beqa}{\begin{eqnarray}}
 
 \providecommand{\rm}{\mathrm}
\providecommand{\eeqa}{\end{eqnarray}}

\newcommand{\beq}{\begin{equation}}
\newcommand{\eeq}{\end{equation}}

\newcommand{\az}{\alpha_0}

\newcommand{\bz}{\beta_0}

\def\T{{\scriptscriptstyle {\rm T}}}
\def\IR{{\scriptscriptstyle {\rm IR}}}
\def\UV{{\scriptscriptstyle {\rm UV}}}
\def\NL{{\scriptscriptstyle {\rm NL}}}
\def\cutoff{{\scriptscriptstyle {\rm cutoff}}}
\def\46{{\scriptscriptstyle {\rm 4-6}}}
\def\24{{\scriptscriptstyle {\rm 2-4}}}

\title{Extended Effective Field Theory of Inflation}

\author[a,b]{Amjad Ashoorioon,}
\author[a,c]{Roberto Casadio,}
\author[a,c,d]{Michele Cicoli,}
\author[e,f]{Ghazal Geshnizjani,}
\author[e]{Hyung J. Kim}

\affiliation[a]{\small INFN, Sezione di Bologna, viale Berti Pichat 6/2, 40127 Bologna, Italy}
\affiliation[b]{\small School of Physics, Institute for Research in Fundamental Sciences (IPM),\\
P.O.Box 19395-5531, Tehran, Iran}
\affiliation[c]{\small Dipartimento di Fisica e Astronomia, Universit\`a di Bologna, \\ via Irnerio 46, 40126 Bologna, Italy}
\affiliation[d]{\small Abdus Salam ICTP, Strada Costiera 11, Trieste 34151, Italy}
\affiliation[e]{Department of Applied Mathematics, University of Waterloo\\ Waterloo, Ontario, N2L 3G1, Canada}
\affiliation[f]{Perimeter Institute for Theoretical Physics\\ 31 Caroline St. N., Waterloo, ON, N2L 2Y5, Canada}

\emailAdd{amjad.ashoorioon@bo.infn.it}
\emailAdd{roberto.casadio@bo.infn.it}
\emailAdd{mcicoli@ictp.it}
\emailAdd{ggeshniz@uwaterloo.ca}
\emailAdd{h268kim@uwaterloo.ca}

\abstract{We present a general framework where the effective field theory of single field inflation is extended by the inclusion of operators with mass dimension 3 and 4 in the unitary gauge. These higher dimensional operators introduce quartic and sextic corrections to the dispersion relation. We study the regime of validity of this extended effective field theory of inflation and the effect of these 
higher dimensional operators on CMB observables associated with scalar perturbations, such as the speed of sound, the amplitude of the power spectrum and the tensor-to-scalar ratio. Tensor perturbations remain instead, unaltered.}

\arxivnumber{1802.03040}
\subheader{IPM/P-2018/006}

\begin{document}
\maketitle

\section{Introduction}
\label{Intro}

The Effective Field Theory of Inflation (EFToI)~\cite{Cheung:2007st} provides an alternative picture to understand perturbations in single field driven inflationary models. In this picture, one first fixes the gauge by taking constant time hypersurfaces to coincide with inflaton constant surfaces. In other words, the perturbations of the inflaton field are absorbed into the fluctuations of the metric. The inflaton fluctuations transform non-linearly under the time diffeomorphism but the non-linear sigma model describing it can always be UV completed into the theory of inflation with a linear representation of time diffeomorphism. The Goldstone mode, $\pi$, corresponding to breaking time diffeomorphism symmetry, describes the scalar perturbations around the FRW background. Therefore, the most generic action for inflation can be constructed out of the quantities that respect the remaining symmetries, {\it i.e.}~the spatial diffeomorphisms. The coefficients of these terms can be time-dependent in principle. 

This means including terms such as the metric component $g^{00}$ and the extrinsic curvature of constant time hypersurfaces, $K_{\mu\nu}$, which are only spatially diffeomorphism invariant, in addition to standard 4-diffeomorphism invariant terms. It was shown in~\cite{Cheung:2007st} that, around a FRW background, this reduces to including functions of time, $g^{00}$ and perturbations of the extrinsic curvature, $\delta K_{\mu\nu}$. 

One can then make the Goldstone mode explicit by applying the St\"uckelberg technique and study the physics of the Goldstone mode at energies where the mixing with gravitons can be neglected. In this framework, the standard slow-roll inflationary model corresponds to adding only operators with time dependence and at most linear in terms of $g^{00}$ (mass dimension zero), before performing the St\"uckelberg trick,. Higher dimensional operators quantify deviations from the slow-roll inflation. On the other hand, in terms of $\pi$ and in the decoupling limit, the slow-roll Lagrangian will include dimension 4 operators from the kinetic terms of the canonically normalised $\pi$. With higher order or mass dimensional operators in the unitary gauge, dimension 6 operators from cubic interactions of the Goldstone modes are generated. Due to the non-linear nature of the time diffeomorphism, once it is restored, the spatially diffeomorphism invariant operators will lead to a non-linear dependence on the Goldstone mode. This in turn leads to non-vanishing higher point correlation functions. Therefore, the coefficients of these operators can be constrained by measuring the corresponding correlation functions.

The above approach is borrowed from particle physics since also the Standard Model is described through the lowest dimension operators that are compatible with the symmetries of the system. Then higher dimensional operators describe deviations from the vanilla case and quantify the emergence of new physics. Similarly, in the EFToI, higher dimensional operators can reproduce various interesting inflationary models, such as DBI inflation \cite{Alishahiha:2004eh, Shiu:2006kj, Shandera:2006ax}, with speed of sound different from one, or Ghost Inflation \cite{ArkaniHamed:2003uz}, with leading spatial gradient term proportional to $(\nabla^2 \pi)^2$\cite{Cheung:2007st}. In particular, as shown in \cite{Cheung:2007st}, reducing the speed of sound naturally strengthens the cubic interaction, which in turn enhances the non-gaussianity.

In the EFToI, these higher dimensional operators can also change the dispersion relation. For example, the inclusion of $\delta K_{\mu\nu}^2$ operators modifies the dispersion relation to $\omega^2\sim k^4$ (similar to Ghost Inflation). Including additional higher dimensional operators of the form $(\nabla^{(n-2)} K_{\mu\nu})^2$ with $n\geq 3$, when they are dominant, would modify the dispersion relation even further to $\omega^2\sim k^{2n}$. 

However, the authors of \cite{Cheung:2007st} did not include operators beyond mass dimension 2, since they claimed that these new terms would not be compatible with a sensible EFT description. The reason is the fact that in such theories, when $E\rightarrow s\, E$, the energy scaling dimension for $\pi$ would be $\pi\to s^{-1/2+3/2n}\,\pi$ and so the scaling behaviour of cubic operators like $\dot{\pi} (\nabla\pi)^2$ would be $(7-3n)/(2n)$. This would imply that for $n\geq 3$ this operator would become \textit{relevant}, whereas for $n<3$ it is \textit{irrelevant} at low energies. Hence a theory with a dispersion relation of the form $\omega^2\sim k^6$ would have an IR strong coupling cut-off $\Lambda_6^\IR$ which would in general make this theory not a controllable EFT. 

However, we argue here that this may not always be the case. Our reason is that a dispersion relation of the form $\omega^2\sim k^6$ would not hold up all the way to low energy scales. It is true that the scaling power of $\dot{\pi} (\nabla\pi)^2$  is $-1/3$, but the scaling power of $(\nabla^2\pi)^2$ and $(\nabla\pi)^2$ are respectively $-2/3$ and $-4/3$. This implies that the dispersion relation would also change from $\omega^2\sim k^6$ to $\omega^2\sim k^4$ at some energy scale $\Lambda_{\rm dis}^\46$. Hence, if the coefficients of the higher dimensional operators are tuned such that $\Lambda_6^\IR \ll \Lambda_{\rm dis}^\46$, the $\omega^2\sim k^6$ theory remains weakly coupled throughout its evolution down to low energies. On top of this, one has to require that the scale $\Lambda_4$, where the $\omega^2\sim k^4$ theory becomes strongly coupled, is not below $\Lambda_{\rm dis}^\46$. This is guaranteed if $\Lambda_6^\IR \ll \Lambda_{\rm dis}^\46 \ll \Lambda_4$. Similarly, one has to require that the standard $\omega^2\sim k^2$ theory does not become strongly coupled at a scale $\Lambda_2$ which is below the scale $\Lambda_{\rm dis}^\24$ where the dispersion relation becomes dominated by the quartic term: $\Lambda_{\rm dis}^\24 \ll \Lambda_2$ \cite{Baumann:2011su}.

Under these conditions, the EFT for the perturbations remains weakly coupled all the way from low energy to the $\omega^2\sim k^6$ regime. Note that in addition, this framework is only valid on energy scales below $\Lambda_b$, where time diffeomorphisms get spontaneously broken by the inflaton background. Thus, we also need to require $\Lambda_{\rm dis}^\46 \ll \Lambda_b$. This is because at $\Lambda_b$ the scalar perturbations become of the same order of magnitude of the background which, therefore, cannot be integrated out to leave an EFT for the fluctuations only. The cut-off of the EFT for the fluctuations $\Lambda_\cutoff$ is therefore the minimum between $\Lambda_b$ and the UV strong coupling scale of the $\omega^2\sim k^6$ theory,  $\Lambda_6^\UV$, associated with some higher derivative operators which remain irrelevant. 

Strictly speaking, the cosmological experiment is done at energies of order the Hubble scale $H$ where horizon exit takes place. Hence, in order for the EFT approach to be under control, one would need just to require $H\ll \Lambda_\cutoff$ and a horizon crossing that occurs in the region where the dispersion relation takes the standard quadratic form. However, the goal of this paper is to provide a consistent theoretical framework to motivate the emergence of modified dispersion relations which, as previously shown, can lead to (super-excited) non-Bunch-Davis initial conditions with interesting implications for CMB observables \cite{Ashoorioon:2017toq}. Thus we need to control the EFT also in the high energy $\omega^2 \sim k^6$ regime which describes the behavior of the perturbations deep inside the horizon. These perturbations are born in the vacuum of $k^6$ theory but then become excited states due to the transitions to the quartic and finally to the standard quadratic regime for the dispersion relation which happen before horizon crossing. Therefore, in order to be able to describe consistently the whole evolution of these modes from deep inside the horizon to super-horizon scales, we need to be able to trust the EFT for the fluctuations up to the $\omega^2 \sim k^6$ regime. 

In this paper we shall therefore include higher order corrections which in the unitary gauge correspond to operators with mass dimension 3 and 4 that were so far neglected in the literature. Hence we shall call our framework Extended Effective Field Theory of Inflation (EEFToI). As explained above, these higher dimensional terms modify the dispersion relation and, depending on the sign and magnitude of their coefficients, can have important implications for scalar perturbations. We shall focus only on modifications of the equation of motion for the Goldstone boson $\pi$ by higher order terms, and so we shall consider operators which include at least one factor of $\delta K_{\mu\nu}$ in their expression. One can write down other operators with the same mass dimensions as the one we focus on below, which would not affect the two point function for scalar perturbations. We also drop operators of the form $(1+g^{00})^n$ with $n\geq 3$ since, contrary to $(1+g^{00})^2$, they do not change the speed of sound for $\pi$.

The outline of this paper is as follows. In Sec.~\ref{EEFT}, we first briefly review the formalism of the EFToI and then we present the EEFToI by writing down all operators with mass dimension three and four that can add a correction proportional to $k^6$ in the dispersion relation. In Sec.~\ref{SecImpl}, we discuss the implications of these higher dimensional operators for tensor and scalar perturbations. In particular, after identifying the allowed operators, we focus on the healthy region of parameter space for $\pi$ and explore various scenarios that can arise.  We study the scalar power spectrum in these scenarios, depending on the sign of the coefficients of the quartic and sextic corrections to the dispersion relation. We also analyse the implications for the tensor-to-scalar ratio. In Sec.~\ref{NonG}, we comment on the cut-off of the EEFToI and the size of non-gaussianity, while in Sec.~\ref{Concl}, we conclude and give directions for future research.

\section{Extended EFT of inflation}
\label{EEFT}

\subsection{Overview of the EFT of inflation}

We start by reviewing the EFToI framework which allows one to write down the most generic action for single scalar field models on a quasi de~Sitter background. The idea is that even though the action is invariant under all diffeomorphisms (4-diffs), the background solution provides a preferred time foliation. By restricting time hypersurfaces to constant inflaton surfaces (the unitary gauge), the action has to satisfy only 3-dimensional spatial diffeomorphism (3-diff) invariance. Therefore, one can start by writing the most general 3-diff invariant action around the FRW metric and the effect of different terms. This implies, in addition to standard 4-diff invariant metric terms, including $g^{00}$ and pure functions of time $f(t)$ that are also scalars under 3-diffs. We can also include terms depending on the extrinsic curvature of the constant time hypersurfaces, $K_{\mu\nu}$, as this is a tensor under spatial diffs. It turns out that any additional 3-diff invariant object, such as the 3-dimensional covariant derivative, can be rewritten in terms of the quantities already included.
It can be shown that the resulting Lagrangian around flat FRW reduces to \cite{Cheung:2007st}
\beq 
\mathcal{L}
=
M_{\rm Pl}^2
\left[
\frac12\, R
+\dot H\,  g^{00}
-\left(3\, H^2 +\dot H\right)
\right]
+\sum_{m\geq 2}\,
\mathcal{L}_m (g^{00}+1,\delta K_{\mu\nu}, \delta R_{\mu\nu\rho\sigma}, \nabla_\mu;t)
\ ,
\label{genspacei}
\eeq
where $\mathcal{L}_m$ represent functions of order $m$ in $g^{00}+1$, $\delta K_{\mu\nu}$ and $\delta R_{\mu\nu\rho\sigma}$. Furthermore, the pure time dependent term and coefficient of $g^{00}$ have become functions of $H$ and $\dot{H}$ by satisfying the Friedmann equations in the FRW limit.\footnote{In the original EFToI formalism~\cite{Cheung:2007st}, only the quadratic operators built out of $\delta K_{\mu\nu}=K_{\mu\nu}-K_{\mu\nu}^{\rm FRW}$ were studied. As we will see, dropping higher order operators is not always necessary and one can extend the EFToI correspondingly.} 

Once the generic action in unitary gauge has been written, the gauge condition can be relaxed by allowing the time transformation $t\to t+\xi^0(x^\mu)$. Since the action is no more restricted to a particular time slicing,  time diffeomorphism invariance has to be restored again. This can be achieved by substituting  $\xi_0(x^\mu)$ with a field $-\pi(x^\mu)$ and requiring that it shifts as $\pi(x^\mu)\rightarrow \pi(x^\mu)-\xi_0(x^\mu)$ under time diffeomorphisms. 
Note that when we perform $t\to t+\xi^0(x^\mu)$, we no longer expect the perturbations in inflaton field, $\phi$,  to be zero. In fact, by introducing the Goldstone boson we are representing these perturbations, since  $\delta\phi=-\dot{\phi}_0\xi^0$ and therefore $\pi=\delta\phi/\dot{\phi}_0$. 

This action is very complicated in general. However, the advantage of using this approach is that, when the generalised slow roll approximation is valid, one can ignore all the metric perturbations in the action. Basically, if the typical scale of the time dependence of the coefficients in the unitary gauge is much longer than the Hubble time, the theory effectively decouples from gravity and it is possible to compute the power spectrum neglecting metric perturbations in the action. 

For instance, if we implement this procedure for the action~\eqref{genspacei}, assuming $\mathcal{L}_m=0$ $\forall\, m$, we obtain
\beq
\mathcal{L}_{\rm slow-roll}= -M_{\rm Pl}^2\,  \dot{H}\left(\dot{\pi}^2-{(\partial \pi)^2\over a^2}\right). 
\eeq
One can show, using proper gauge transformations, that $\pi$ is related to the conserved quantity $\zeta$ by $\zeta=-H\pi$. Substituting $\zeta$ in the above action reduces it to the standard slow-roll inflationary action for $\zeta$ \footnote{The correspondence is up to a mass correction in higher orders of slow roll parameters. In quasi de Sitter space, $\pi$ gets a small mass due to time-dependent background, which has to be taken into account to get the exact correspondence}. Since we are interested in deviations from the standard slow-roll model, we will turn on the coefficients of the operators in $\mathcal{L}_m$. In the EFToI, focusing only on the terms that contribute to the quadratic action of $\pi$ and can change the dispersion relation up to quartic order, the unitary gauge action is
\beq
\mathcal{L}_{\rm EFToI} = \mathcal{L}_{\rm slow-roll} + \mathcal{L}_2\,,
\label{LEFT}
\eeq
with:
\beq
\label{eftoiunigauge}
\mathcal{L}_2 =
\frac{M_2^4}{2!}\,(g^{00}+1)^2
+\frac{\bar M_1^3} {2}(g^{00}+1) \delta K^\mu_{\ \mu} -\frac{\bar M_2^2}{2}\, (\delta K^\mu_{\ \mu})^2
-\frac{\bar M_3^2}{2}\, \delta K^\mu_{\ \nu}\,\delta K^\nu_{\ \mu}
\ .
\eeq
The first term, $(g^{00}+1)^2$, is the operator that modifies the speed of sound for scalar perturbations from the speed of light. Noting that $(g^{00}+1)$ has zero mass dimension, powers of $(g^{00}+1)^n$ with $n\geq 3$ could also be included which result to more general K-inflationary models. In this work, without loss of generality, we only include the quadratic term in $(g^{00}+1)$ as it explicitly modifies the speed of sound for scalar perturbations. The mass dimension 1 term, $(g^{00}+1) \,\delta K^\mu_{\ \mu} $, is not symmetric under time reversal, and was already analysed in~\cite{Cheung:2007st,Bartolo:2010bj}. Finally $ \frac{\bar M_2^2}{2}\, (\delta K^\mu_{\ \mu})^2$ and $\frac{\bar M_3^2}{2}\,\delta K^\mu_{\ \nu}\,\delta K^\nu_{\ \mu}$ are the operators that lead to generalized Ghost Inflation with a quartic correction to the dispersion relation~\cite{ArkaniHamed:2003uz, Cheung:2007st}.

\subsection{Lagrangian of the Extended EFT of inflation}

We know proceed to introduce the Lagrangian of the EEFToI. We supplement the unitary gauge action of the EFToI with operators of mass dimension 3 and 4 constructed out of $\delta K^\nu_{\ \mu}$, $\nabla_\mu$ and $g^{00}$ so that the Lagrangian of the EEFToI (again keeping only terms which contribute to quadratic action for $\pi$ ) becomes
\beq
\mathcal{L}_{\rm EEFToI} = \mathcal{L}_{\rm EFToI} + \mathcal{L}_{2, d3}+ \mathcal{L}_{2, d4}\,,
\label{Ltot}
\eeq
with:
\begin{eqnarray}
\label{eq:actiontad}
 \mathcal{L}_{2, d3}+ \mathcal{L}_{2, d4}
&=&
\frac{\bar{M}_4}{2} \nabla^\mu g^{00}\nabla^\nu\delta K_{\mu \nu}-\frac{\delta_1}{2}\,(\nabla_{\mu} \delta K^{\nu\gamma})(\nabla^ {\mu} \delta K_{\nu\gamma})
-\frac{\delta_2}{2}\,(\nabla_{\mu} \delta K^\nu_{\ \nu})^2 
\nonumber\\
&&-\frac{\delta_3}{2} \,(\nabla_{\mu} \delta K^\mu_{\ \nu})(\nabla_{\gamma} \delta K^{\gamma\nu})-\frac{\delta_4}{2} \,\nabla^ {\mu}\delta K_{\nu\mu}\nabla^ {\nu}\delta K_{\sigma}^{\sigma},
\end{eqnarray}
where $M_i$, $\bar M_i$ and $\delta_i$'s are free (time dependent) coefficients and the sign of each term is also a priori free. We remark that the mass dimension 3 operator $\nabla^\mu g^{00}\nabla^\nu\delta K_{\mu \nu}$ is not symmetric under time reversal and $\delta_i$'s are the coefficients of dimension 4 operators.

Our motivation for adding such operators is that they lead to sixth order corrections to the dispersion relation. The reason advocated in~\cite{Cheung:2007st} for discarding these operators is that sixth order corrections to the dispersion relation would make higher derivative operators relevant, signaling the presence of an IR strong coupling regime. However, as explained in Sec.~\ref{Intro} and as we will discuss in more depth in Sec.~\ref{NonG}, one can still have a sensible EFT description if the IR strong coupling scale of the $\omega^2\sim k^6$ theory is below the scale $\Lambda_{\rm dis}^\46$ where the dispersion relation becomes dominated by the quartic term. 

We emphasise here that in this EFT approach to inflation, the Lorentz symmetry is not preserved. In particular, the mass dimension and energy scaling dimension of the operators depending on $\delta K$ do not match. We should also add that we have only kept terms that can modify the action for scalar perturbations at second order. There are other operators, e.g.~proportional to $\bar{M}\, {\delta K}^3$,  that do not contribute to the linear equation of motion of the Goldstone boson $\pi$. For reasons that will become clear momentarily, we first focus on the implications of the mass dimension 3 and 4 operators on the action for tensor perturbations. Then we will get back to the action for the Goldstone mode $\pi$ derived from the unitary gauge action in the EEFToI. 

\section{Perturbations in the extended EFT of inflation}
\label{SecImpl}

\subsection{Tensor perturbations}

To study the tensor perturbations, we perturb the spatial part of the metric as follows
\beq
g_{ij}=a^2 (\delta_{ij}+\gamma_{ij})\, ,
\eeq
where $\gamma_{ij}$ is transverse and traceless (summation over the repeated index $i$ is assumed),
\beq\label{tt-cond}
\gamma_{ii}=0\,,\qquad\quad \partial_{i}\gamma_{ij}=0\ ,
\eeq
and we will expand the action~\eqref{Ltot} up to second order in $\gamma_{ij}$. As pointed out in the original EFToI paper~\cite{Cheung:2007st}, from the mass dimension zero up to mass squared operators, only $-\frac{\bar M_3^2}{2}\, \delta K^\mu_{\ \nu}\,\delta K^\nu_{\ \mu}$ affects the equation of motion for $\gamma_{ij}$, and adds the following contribution to the action for tensor perturbations
\beq
S^{\bar M_3^2}=-\frac{\bar M_3^2}{8}\int a^3 dt\, d^3 x ~\partial_0\gamma_{ij}\partial_0\gamma_{ij}
\ .
\label{SbarM32}
\eeq
The action~\eqref{SbarM32}, together with the terms describing the usual massless scalar field contribution for tensor perturbations,
\beq
S_{\rm EH}=\frac{M_{\rm Pl}^2}{8} \int dt\, d^3 x \,a^3 \left(\partial_0\gamma_{ij}\partial_0\gamma_{ij}-a^{-2}\partial_{l}\gamma_{ij}\partial_{l}\gamma_{ij}\right)
\ ,
\eeq
produces the following modified equation of motion for tensor perturbations, 
\beq\label{tensoreq}
p_k''+\left(c_\T^2 k^2 -\frac{a''}{a}\right)p_k=0\,,
\eeq
where $p_k\equiv a\, \gamma_k$, the subscript $k$ refers to the Fourier transform of the two helicities of the gravitons, and 
\beq\label{M3bar-negative-sign}
c_\T^2=\left(1-\frac{\bar M_3^2}{M_{\rm Pl}^2}\right)^{-1}
\ .
\eeq
In order to avoid superluminal propagation for tensor perturbations, one must then have $\bar M_3^2\leq 0$\footnote{ See \cite{Cai:2015yza} for scenarios where  $\bar{M}_3$ and consequently $c_T$ are time dependent.}. Larger values of $|\bar M_3|$ correspond to smaller speed of sound and, for $|\bar M_3|\rightarrow M_{\rm Pl}$, the speed of gravitational waves $c_\T\to {1}/{\sqrt{2}}$. 

For $\bar M_3^2> 0$, the speed of gravitation waves is superluminal. Lorentz invariant EFTs with this property have been argued to be non-local and not embeddable in a local quantum field theory or string theory~\cite{Adams:2006sv}. However, \cite{Babichev:2007dw} argues that such models, despite having superluminal propagation, do not lead to any violation of causality. If $\bar M_3^2> 0$ is taken as a legitimate choice, larger values of $\bar M_3^2$  increase the value of  speed of sound and in fact this value diverges for $\bar M_3^2\to M_{\rm Pl}^2$.

Now let us focus on the new operators with mass dimension 3 and 4. The term $\frac{\delta_1}{2}\,(\nabla_{\mu} \delta K^{\nu\gamma})(\nabla^ {\mu} \delta K_{\nu\gamma})$, up to second order in $\gamma_{ij}$, yields
\beq\label{D1}
\frac{\delta_1}{2} \left[-\frac{H^2}{2} \left(\partial_0 \gamma_{ij}\right)^2+\frac{1}{4 a^2 } \left(\partial^2_{0m}\gamma_{ij} \right)^2-\frac{1}{4}\left(\partial^2_0\gamma_{ij}\right)^2\right]
\ ,
\eeq
where $i$, $j$ and $m$ are spatial indices and summation over $m$ is assumed. Other operators do not contribute to the Lagrangian of tensor perturbations once the transverse traceless condition on $\gamma_{ij}$ is imposed. The appearance of $\partial_0^2 \gamma_{ij}$ in the Lagrangian density~\eqref{D1} leads to ghost instability for the tensor perturbations. Thus the requirement of having no ghosts in the tensor sector of the theory translates into
\beq
\delta_1=0
\ .
\eeq
Therefore, dispersion relation of the primordial tensor fluctuations remains resilient to changes in the scalar and gravitational sector in the EEFToI, like it was shown originally in~\cite{Creminelli:2014wna}.

\subsection{Scalar perturbations}

The Goldstone boson can explicitly appear in the action via the St\"uckelberg trick. In this procedure, the variation of the metric with broken time diffeomorphism, $t\to \tilde{t}=t+\xi^{0}(x)$, $x\to \vec{\tilde{x}}=\vec{x}$, is obtained and $\xi^0(x(\tilde{x}))$ is everywhere replaced by $-\tilde{\pi}(\tilde{x})$ in the transformed action. Then $\pi(x)$ is assumed to transform non-linearly to $\tilde{\pi}(\tilde{x})=\pi(x)-\xi^0(x)$, which guarantees that the action remains invariant under diffs to all orders~\cite{Cheung:2007st}. Evaluating the action explicitly for $\pi$ in Fourier space, Ref.~\cite{Bartolo:2010bj} found out that, in the decoupling limit corresponding to $\dot{H}\to 0$ while $M_p^2\dot{H}$ is fixed, the EFToI Lagrangian~\eqref{LEFT} in the unitary gauge leads to the following second order Lagrangian in the $\pi$-gauge:
\begin{eqnarray}
\mathcal{L}_{\rm EFToI}^{(\pi)}&=& M_p^2 \dot{H} (\partial_{\mu}\pi)^2+2 M_2^4 \dot{\pi}^2-\bar M_1^3 H\left(3\dot{\pi}^2-\frac{(\partial_i \pi)^2}{2 a^2}\right)-\frac{\bar M_2^2}{2} \left(9H^2 \dot{\pi}^2-3 H^2 \frac{(\partial_i \pi)^2}{a^2}\right.\nonumber\\&+&\left.\frac{(\partial_i^2 \pi)^2}{a^4} \right)
-\frac{\bar M_3^2}{2} \left(3H^2 \dot{\pi}^2-H^2 \frac{(\partial_i \pi)^2}{a^2}+\frac{(\partial_j^2 \pi)^2}{a^4}\right)\ ,
\label{Lpi}
\end{eqnarray}
whereas, following the same procedure, the mass dimension 3 parity-violating operator in (\ref{eq:actiontad}) in the same limit $\dot{H}\to 0$ yields
\begin{eqnarray}
\mathcal{L}_{2,\,d_3}^{(\pi)} = \frac{\bar M_4}{2}\left(\frac{k^4 H \pi^2}{a^4}+\frac{k^2 H^3 \pi^2}{a^2}-9 H^3\dot\pi^2\right)\,.
\label{Ld3}
\end{eqnarray} 
This operator only yields a quartic correction to the dispersion relation. On the other hand, the mass dimension 4 operators in the EEFToI action (\ref{Ltot}) give~\footnote{In \cite{Ashoorioon:2017toq}, in the definition for $\delta K_{\mu\nu}\equiv K_{\mu\nu}-K^{(0)}_{\mu\nu}$, we had mistakenly subtracted the FRW result $K^{(0)}_{\mu\nu}=-a^2 H\delta_{ij}\delta_\mu^i\delta_\nu^j$ from the extrinsic curvature of constant $t$ spatial surfaces. However, this would not produce a tensorial structure for $\delta K_{\mu\nu}$. Following \cite{Cheung:2007st}, one has to instead subtract $K^{(0)}_{\mu\nu}=-H h_{\mu\nu}$, which reduces to the FRW result when the perturbations vanish and guarantees that $\delta K_{\mu\nu}$ is covariant. The difference in the obtained results for the action of Goldstone boson is due to this. We thank P.~Creminelli for clarification of this issue to us. We take the opportunity to correct the mistake/typo in \cite{Cheung:2007st} in defining $K^{(0)}_{\mu\nu}$ as $a^2 H h_{\mu\nu}$. There seems to be an extra factor of $-a^2$ in their expression. }

\begin{eqnarray}\label{action-pi}
\mathcal{L}_{2,\,d_4}^{(\pi)}&=&-\frac{1}{2} \delta_1 \left(\frac{k^6 \pi ^2}{a^6}-\frac{3 H^2 k^4 \pi ^2}{a^4}-\frac{k^4 \dot{\pi}^2}{a^4}+\frac{4 H^4 k^2 \pi ^2}{a^2}-6 H^4 \dot{\pi}^2-3 H^2 \ddot{\pi}^2\right)\nonumber\\&&
-\frac{1}{2} \delta_2 \left(\frac{k^6 \pi ^2}{a^6}+\frac{H^2 k^4 \pi ^2}{a^4}-\frac{k^4 \dot{\pi}^2}{a^4}+\frac{6 H^4 k^2 \pi ^2}{a^2}-9 H^2 \ddot{\pi}^2\right)\nonumber\\&&
-\frac{1}{2} \delta_3 \left(\frac{k^6 \pi ^2}{a^6}+\frac{3 H^2 k^4 \pi ^2}{a^4}+\frac{ H^2 k^2 {\dot{\pi}} ^2}{a^2}-9 H^4 \dot{\pi}^2\right)\nonumber\\&& -\frac{1}{2}\delta_4 \left(\frac{k^6 \pi ^2}{a^6}+\frac{ H^2 k^4 \pi ^2}{2 a^4}+\frac{9 H^4 k^2 \pi ^2}{2 a^2}+\frac{3 H^2 k^2 \dot{\pi}^2}{a^2}+\frac{27}{2} H^4 \dot{\pi}^2\right)\,.\nonumber\\
\end{eqnarray}
The terms $-\frac{\delta_1}{2}(\nabla_{\mu} \delta K^{\nu\gamma})(\nabla^{\mu} \delta K_{\nu\gamma})$ and $-\frac{\delta_2}{2}(\nabla_{\mu} \delta K^\nu_{\ \nu})^2 $ produce $\ddot{\pi}^2$. According to Ostrogradski's theorem~\cite{Ostrogradski}, higher time derivatives usually lead to ghost instabilities. In principle such a ghost term could be avoided if $\delta_1=-3\,\delta_2$. However, as noticed in the previous section, one has to set $\delta_1=0$ in order to avoid ghosts in the tensor sector, which enforces 
\beq
\delta_1=\delta_2=0
\ .
\eeq
Therefore, only the terms $\frac{\delta_3}{2}\nabla_{\mu} \delta K^\mu_{\ \nu} \nabla_{\gamma} \delta K^{\gamma\nu}$ and $\frac{\delta_4}{2} \,\nabla^ {\mu}\delta K_{\nu\mu}\nabla^ {\nu}\delta K_{\sigma}^{\sigma}$ are allowed. 

We shall now derive the equation of motion for $\pi$ using the total EEFToI Lagrangian in the $\pi$ language and in the decoupling limit, which is the sum of (\ref{Lpi}), (\ref{Ld3}) and (\ref{action-pi}), 
\beq
\mathcal{L}_{\rm EEFToI}^{(\pi)} = \mathcal{L}_{\rm EFToI}^{(\pi)} + \mathcal{L}_{2,\,d_3}^{(\pi)} + \mathcal{L}_{2,\,d_4}^{(\pi)}
\label{Lpitot}
\eeq
Before that, we briefly discuss the equation of motion corresponding to the EFToI Lagrangian. Variation of $\mathcal{L}_{\rm EFToI}^{(\pi)}$, yields the same equation of motion previously derived in \cite{Bartolo:2010bj},
\beq\label{pi-eq1}
A_0\,\ddot{\pi}_k +(B_0+3H C_0)\,\dot{\pi}_k+\left(E_0 \frac{k^4}{a^4}+D_0\frac{k^2}{a^2}\right)\pi_k=0
\eeq
with
\beqa
A_0&=&-2 M_{\rm Pl}^2 \dot{H}+4 M_2^4-6\bar M_1^3 H-9 H^2 \bar M_2^2-3 H^2 \bar M_3^2 \nonumber\\
B_0&=& -6 \bar M_1^3 \dot{H}-18 \dot{H}{H}\bar M_2^2 -6\dot{H} H \bar M_3^2\nonumber\\
C_0&=&-2 M_{\rm Pl}^2 \dot{H}+4 M_2^4-6\bar M_1^3 H-9 H^2 \bar M_2^2 -3H^2 \bar M_3^2 \nonumber\\
D_0 &=&-2 M_{\rm Pl}^2 \dot{H}-\bar M_1^3 H-3H^2 \bar M_2^2-\bar M_3^2 H^2\nonumber\\
E_0&=&\bar M_2^2+\bar M_3^2
\ ,
\eeqa
where the mixing with gravity has been neglected. Noting that the canonical $\pi_c\sim \sqrt{A_0}\, \pi$ and $\delta g^{00}_{c}\sim M_{\rm Pl}\, \delta g^{00}$, the mixing energy between gravity and the Goldstone mode $\pi$ can be neglected at energies
\beq
E> E_{\rm mix}\sim \frac{\sqrt{A_0}}{M_{\rm Pl}}
\ ,
\eeq
which is known as the equivalence theorem. Namely, the perturbations of the longitudinal mode decouple from the perturbations of the metric at energies higher than $E_{\rm mix}$. Being able to compute the correlation functions at horizon crossing corresponds to assuming $H > E_{\rm mix}$. In the regime where $M_{\rm Pl}^2 \dot{H}$ is bigger than the other contributions to $A_0$, $E_{\rm mix}\sim \epsilon^{1/2}H$, and so the decoupling limit is guaranteed by the slow-roll condition $\epsilon\ll 1$. On the other hand if $M_2^4$ is bigger than the other parameters in $A_0$, $E_{\rm mix}\sim M_2^2/M_{\rm Pl}$, and decoupling happens if $M_2^2<M_{\rm Pl}\, H$. These conditions were obtained in \cite{Cheung:2007st} too. Other limiting cases suggest $\bar{M}_1^{3/2}<M_{\rm Pl}\,H^{1/2}$, $\bar M_2<M_{\rm Pl}$ and $\bar M_3<M_{\rm Pl}$.

In terms of the conformal time, $d\tau\equiv {dt}/{a}$ and through a change of variable $u_k= a\pi_k$, the equation of motion \eqref{pi-eq1} takes the following form 
\beqa
u_k''+\frac{B_0+3 H C_0-3 A_0 H}{A_0}\, a u_k'+\left(\frac{D_0}{A_0}k^2+\frac{E_0}{A_0}{k^4}{a^2}-\frac{a''}{a}\right)u_k=0\,.\nonumber \\
\eeqa
Noting that $B_0+3 H C_0-3 A_0 H=-6 \dot{H}(\bar{M}_1^3+3H^2 \bar{M}_2+H \bar{M}_3^2 )$, in the limit $\dot{H}\rightarrow0$, this equation is reduced to
\beq
u_k''+\left(\frac{D_0}{A_0}k^2+\frac{E_0}{A_0}{k^4}{a^2}-\frac{a''}{a}\right)u_k=0\,.
\eeq
which has the same form derived in \cite{Bartolo:2010bj}.

If we now include the higher dimensional corrections $\mathcal{L}_{4,\,d_3}^{(\pi)}$ and $\mathcal{L}_{4,\,d_4}^{(\pi)}$ to the total Lagrangian (\ref{Lpitot}), the equation of motion for $\pi$ becomes\footnote{See \cite{Cai:2016thi} for a different way of deriving sixth order dispersion relations. }
\beq\label{pi-eom-total}
A_1 \,\ddot{\pi}_k+B_1\,\dot{\pi}_k+\left(C_1 \frac{k^6}{a^6}+D_1\frac{k^4}{a^4}+F_1 \frac{k^2}{a^2}\right)\pi_k=0\,.
\eeq
Defining
\beq
F_0(k,\tau)=\frac{9}{2}\delta_3-\frac{27}{4}\delta_4- \frac{k^2}{2 a^2 H^2}(\delta_3+3\delta_4)-\frac{9}{2H}\bar M_4
\ ,
\eeq
the coefficients of the equation of motion \eqref{pi-eom-total} read
\beqa
&&A_1=A_0+2 H^4 F_0(k,\tau)\,,\nonumber\\
&&B_1= B_0+3H C_0+H^5\left[6 F_0(k,\tau)+\frac{2 k^2}{a^2 H^2}\left(\delta_3+3\delta_4\right)\right]\,,\nonumber\\
&&C_1=\delta_3+\delta_4\,,\nonumber\\
&&D_1=  E_0+H^2 \frac{\delta_4}{2}+3H^2 \delta_3-\bar M_4 H\,,\nonumber\\
&& F_1=D_0+3 H^4 \left(\delta_3+\frac{3}{2}\delta_4\right)-\bar{M}_4 H^3
\ ,
\eeqa
where we have assumed that the time-dependence of the coefficients is slow compared to the Hubble time. This means that the terms coming from the Taylor expansion of the coefficients are small. 

Expressing the equation of motion for $u_k=a\,\pi_k$, in conformal time and again in the limit $\dot{H}\to 0$, we obtain 
\beqa
&&u_k''+ \frac{2 k^2  H^3 (\delta_3+3 \delta_4 )}{a A_1}\, u_k'+u_k \left(\frac{C_1}{A_1}\frac{k^6}{a^4}+\frac{D_1}{A_1}\frac{k^4}{a^2}+\frac{F_1}{A_1}k^2-\frac{a''}{a}\right)=0\,.
\eeqa
In a de Sitter space, where $a=-1/(H\tau)$, the above equation reads
\beqa\label{equ}
u_k''&+& \frac{G_3 \tau k^2 }{G_1+G_2 \tau^2 k^2}\,u_k'+u_k 
\left(\frac{F_2}{G_1+G_2 \tau^2 k^2 }k^2+\frac{D_2 k^2}{G_1+G_2 k^2 \tau^2}k^2 \tau^2\right.\nonumber\\&+&\left.\frac{C_2 k^2}{G_1+G_2\tau^2 k^2} k^4\tau^4-\frac{2}{\tau^2} \right)=0
\eeqa
where 
\beqa\label{var2}
G_1&\equiv&A_0+9 H^4 \left(\delta_3-\frac{3}{2}\delta_4\right)-9\bar M_4 H^3\nonumber\\
G_2&\equiv& - H^4 \left(\delta_3+3\delta_4\right)\nonumber\\
G_3&\equiv&- 2 H^4 \left(\delta_3+3\delta_4\right)=2 G_2\nonumber\\
F_2&\equiv&F_1 \nonumber\\
D_2&\equiv&D_1 H^2 \nonumber\\
C_2&\equiv&C_1 H^4
\eeqa
and 
\beq
A_1(k,\tau)=G_1+G_2 k^2 \tau^2
\ .
\eeq
The speed of propagation of the perturbations in the intermediate IR, i.e. for $1 < \kappa \tau < \sqrt{G_1/G_2}$, can now be read out from eq.~\eqref{equ} to be
\beq
c_s^2 =\frac{F_2}{G_1}
\ .
\label{cs2FG}
\eeq
In terms of a dimension less variable, $x\equiv k\tau$, the equation of motion \eqref{equ} takes the form
\beq
\frac{d^2 u_k}{d x^2}+\frac{G_3 x}{G_1+G_2 x^2}\frac{d u_k}{d x}+\left(\frac{F_2}{G_1+G_2 x^2}+\frac{D_2 x^2}{G_1+G_2 x^2}+\frac{C_2 x^4}{G_1+G_2 x^2}-\frac{2}{x^2}\right) u_k=0
\ .
\label{eom}
\eeq
In general, due to non-polynomial scale dependence of the dispersion relation, the sound wave analogy does not necessarily apply. A variety of scenarios can in principle arise, depending on the size and sign of various coefficients. For example, if $G_1$ and $G_2$ are positive, the modes start with $c_s^2\sim 0$ in infinite past and gradually, achieve a constant $c_s^2$ given by $F_2/G_1$ as they finally exit the horizon for $|k\tau|\ll 1$. If one of $G_1$ or $G_2$ (but not both) is negative, there could be a singular time in the evolution of each mode from inside the horizon when the speed of sound becomes infinite. It is also possible that both the speed of sound and the coefficient of the quartic correction to the dispersion relation are negative. These instabilities can in principle be avoided in the UV by a positive sextic correction to the dispersion relation. Even though such scenarios are all interesting, we will not focus on them in this paper. 

In the following, we set $\delta_3=-3\,\delta_4$, so that the equation of motion (\ref{eom}) simplifies to~\footnote{With $\delta_3\neq -3\, \delta_4$, the mixing between gravity and the Goldstone boson becomes time-dependent for each mode. Even though at horizon crossing, one can tune the parameters to make $H>E_{\rm mix}$, when the physical momentum of the mode $k/a>M_{\rm Pl} (\delta_3+3\delta_4)^{-1/2}$, the mixing with gravity cannot be neglected. At such momenta the theory of General Relativity also breaks down and in absence of a quantum theory of gravity the predictions of the theory becomes unreliable.}
\beq\label{mode-eq-x}
\frac{d^2 u_k}{d x^2}+\left(\frac{F_2}{G_1}+\frac{D_2 }{G_1}x^2+\frac{C_2 }{G_1}x^4-\frac{2}{x^2}\right) u_k=0\,.
\eeq
If ${F_2}/{G_1}> 0$, as assumed here, our definition of the speed of sound for scalar perturbations (\ref{cs2FG}) is real and well-defined:
\beq
c_s^2=\frac{F_2}{G_1}
\ .
\eeq
If ${F_2}/{G_1}<0$, one deals with the sort of dispersion relation that comes up in Ghost Inflation-like scenarios~\cite{ArkaniHamed:2003uz,Senatore:2009gt}. This will not be something that we focus on in this investigation. Moreover, in order to avoid potential issues related to superluminal propagation in the IR and a unitary UV completion, we assume that \cite{Adams:2006sv}
\beq
0\leq c_s^2\leq 1
\ .
\eeq
The first inequality could be satisfied if $F_2$ and $G_1$ have the same sign. Note that it is also possible to have $c_s^2=1 $ if
\beq
4 M_2^4-5\bar M_1^3 H-6 H^2 \bar M_2^2-2 H^2 \bar M_3^2-9H^4 \delta_4-8\bar M_4H^3=0\,
\eeq
eventhough, higher-point interactions will not be zero. This shows that, although in the EFToI one tries to quantify the deviations from the standard slow-roll model with the operators $\mathcal{L}_m$ in~\eqref{genspacei}, there is still the possibility that one flows back close to the regime of slow-roll inflation with $c_s=1$ via appropriate tuning of the coefficients of the various higher dimensional operators. This result was also true in the EFToI with $\delta K^2$ terms in the unitary gauge Lagrangian. Although one expects to see an enhancement in the non-gaussianity amplitudes due to the modified dispersion relation.

Defining $x'\equiv c_s\, x$, the equation of motion (\ref{mode-eq-x}) reduces further to
\beq\label{eom-u-xprime}
\frac{d^2 u_k}{dx'^2}+\left(1+\alpha_0 x'^2+\beta_0 x'^4-\frac{2}{x'^2}\right)u_k=0
\ ,
\eeq
where 
\beq\label{ab0}
\alpha_0\equiv \frac{D_2}{G_1 c_s^4}=\frac{D_2 G_1}{F_2^2}\, \qquad \beta_0\equiv\frac{C_2}{G_1 c_s^6}=\frac{C_2 G_1^2}{F_2^3}
\ .
\eeq
In order to have a stable vacuum in the deep UV, where the dispersion relation is dominated by the $k^6$, we consider $\beta_0>0$\footnote{Dispersion relations dominated by the $k^6$ term can also be motivated from studies in the context of condensed matter physics applied to black holes~\cite{Jacobson:1991gr}.}. If Hubble crossing takes place when $\omega^2\sim k^2$, transitions between regions with a different dispersion relation can lead to excited initial states \cite{Ashoorioon:2017toq}. In this case, if $\alpha_0<0$, one can have relatively large corrections, much bigger than one, compared to the standard inflationary predictions for the two point function. This can happen even if the dispersion relation has a single turning point, {\it i.e.}~a single horizon crossing is assumed to occur.\footnote{We are referring to $1+\alpha_0 x'^2+\beta_0 x'^4\sim\frac{2}{x'^2}$ as horizon crossing.} The latter scenario is possible if $D_2/G_1<0$ and $C_2/G_1>0$ \footnote{In solid state physics, the perturbations that appear close to the minimum of the dispersion relations at large $k$'s are known as {\it rotons}, the ones that are close to the maximum of the dispersion relation are called {\it maxons} and the ones that are in the linear regime of the dispersion relation are {\it phonons}. In the cosmological setup, starting from deep inside the horizon, each mode in principle can undergo all three phases until it exits the horizon.}.

One should note that the coefficients $\bar M_i$'s, with $i=1,2,3$, in \eqref{eftoiunigauge} can appear with either sign. In fact, as noticed in the previous section about tensor perturbations, one has to assume $\bar M_3^2<0$ in order to avoid superluminal tensor perturbations. Noting that 
\beq
D_2=(\bar M_2^2+\bar M_3^2+ H^2 \frac{\delta_4}{2}-\bar M_4H)H^2
\ ,
\eeq
using the definition~\eqref{ab0}, the condition $\delta_3=-3 \delta_4$ and the expressions~\eqref{var2} for $C_2$ and $G_1$, it is not difficult to show that
\beq
D_2=-\frac{2\alpha_0}{\beta_0 c_s^2} \delta_4 H^4
\ .
\eeq
As stated before, to achieve the stability in the UV, one has to assume that $\beta_0>0$, and hence, $D_2$ and $\alpha_0 \delta_4$ must have opposite signs. If $\delta_4>0$, noting that $C_2=-2\delta_4 H^4$ will be negative, one has to assume $G_1<0$ to obtain a positive $\beta_0$.  Assuming $\dot{H}<0$, that puts an upper bound on how large the first slow-roll parameter $\epsilon$ can be in this scenario. Then in order to achieve $\alpha_0<0$, one has to assume that $D_2>0$. This combined with that $\bar M_3^2<0$ for tensor perturbations to be subluminal, means
\beq
 |\bar M_3|^2<\bar M_2^2+ \frac{\delta_4}{2} H^2-\bar M_4H\,.
\eeq
On the other hand, if $\delta_4<0$, one has to assume $D_2<0$ and the reverse of the above equality should hold and, conversely, a lower bound on $\epsilon$ is found under the assumption that $\dot{H}<0$. One should emphasize that the value of $\alpha_0$ and $\beta_0$ are not solely determined by $c_s$. There are other variables involved in these parameters and even for the choice $c_s\approx1$, $\alpha_0$ and $\beta_0$  can be substantial. Of course, with $c_s\ll 1 $, the values of these parameters will get larger, but a reduced sound speed is not necessary to have large corrections to the dispersion relation. For $c_s\approx 1$, since non-gaussianities are usually determined by factors $1/c_s^2$, one expects at least not to get an enhancement from speed of sound. Nonetheless the non-linear evolution of the mode inside the horizon due to the non-linear dispersion relations, may enhance the value of the non-gaussianity and bring it potentially to a value that can be observed. In fact, if, as in \cite{Ashoorioon:2017toq}, the evolution of the modes inside the horizon can be mapped into excited states with large occupation numbers at horizon crossing, it is expected that, even for $c_s=1$, equilateral \cite{Agullo:2010ws,Ashoorioon:2013eia} and flattened non-gaussianity configurations \cite{Holman:2007na, Ashoorioon:2010xg} will get enhanced. Whether such an equivalence holds and how higher dimensional operators in the EEFToI contribute to the bispectrum shapes and amplitude is something that we plan to investigate in near future.
 
\begin{figure}[b]
\includegraphics[width=\textwidth]{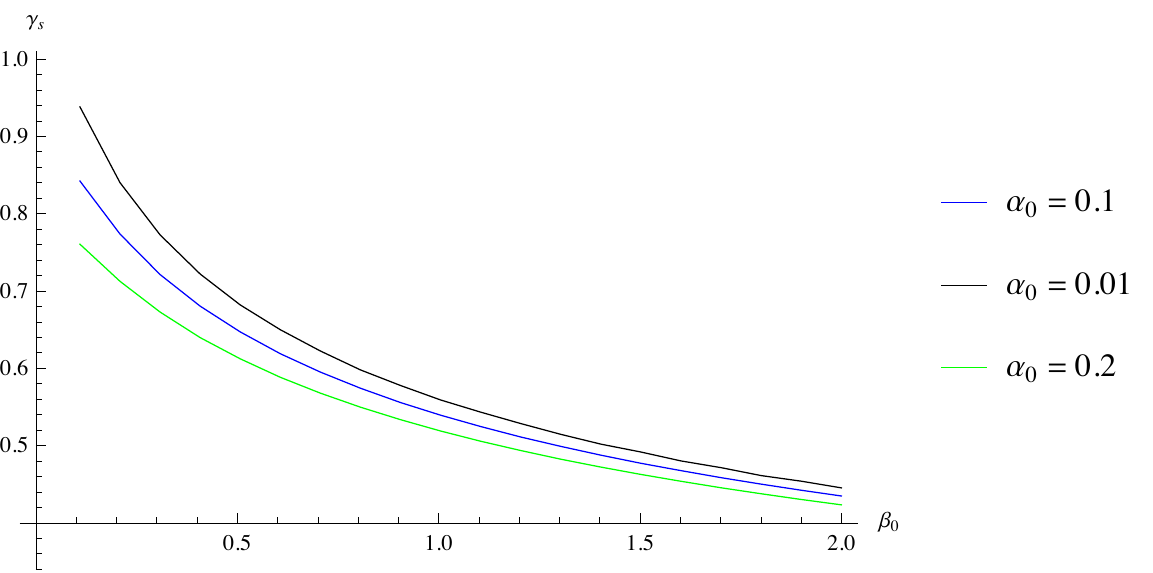}
\caption{Plot of $\gamma_s$  as a function of $\beta_0$ when $\alpha_0$ and $\beta_0$ are both positive and $0.01\leq \alpha_0<1$ with $0\leq \beta_0\leq2$.
\label{a0-b0-pos-1}}
\end{figure}
\begin{figure}[t]
\includegraphics[width=\textwidth]{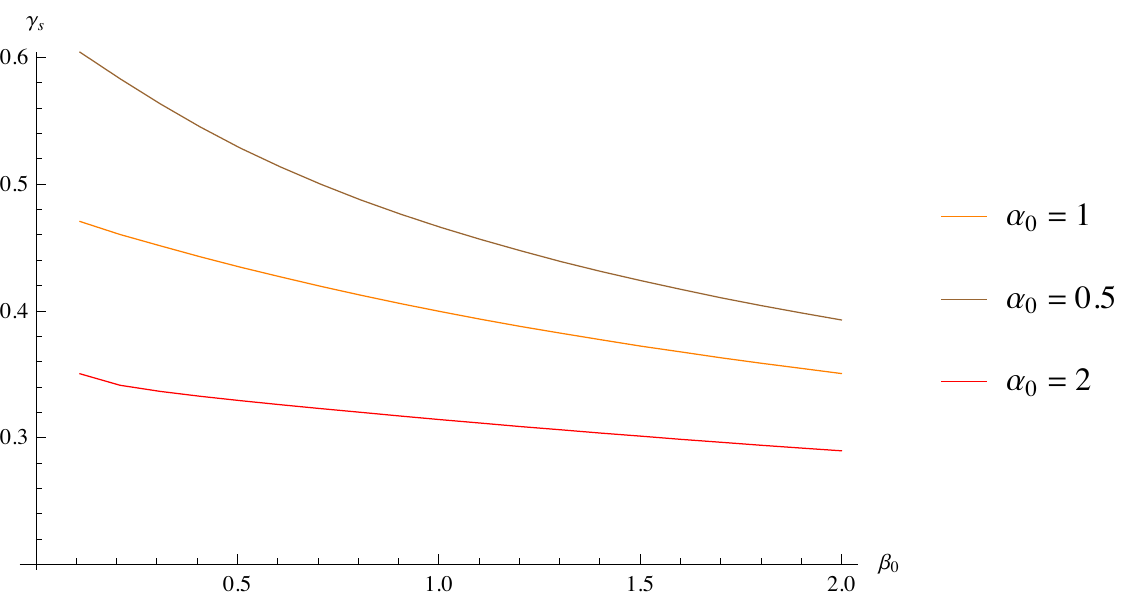}
\caption{Plot of $\gamma_s$  as a function of $\beta_0$ when $\alpha_0$ and $\beta_0$ are both positive and $1\leq \alpha_0\leq2$ with $0\leq \beta_0\leq2$.
\label{a0-b0-pos-2}}
\end{figure}

\subsection{Scalar power spectrum}
\label{Sps}

In this section we investigate the effect of the modified mode equation~\eqref{eom-u-xprime} on the scalar power spectrum for different scenarios, corresponding to different signs and magnitudes of the coefficients of the quartic and sextic corrections to the dispersion relation. In general, overall correction to standard the Bunch-Davies power spectrum can be expressed as a multiplicative factor
\beq
P_s=\gamma_s \,P_s^{\rm B.D.}=\gamma_s \,\frac{H^2}{8\pi^2 c_s \epsilon}
\ .
\eeq
If both $\alpha_0>0$ and $\beta_0>0$, the mode evolution inside the Hubble patch remains oscillatory throughout, and unless the parameters $\alpha_0$ and $\beta_0$ are of order one, one does not expect to see a significant modification to the Bunch-Davies power spectrum. We solve the mode equation~\eqref{eom-u-xprime} numerically, from the positive frequency WKB mode in the infinite past,
\beq
\label{initcond2}
u_k(x'\to-\infty)
\simeq
\frac{1}{2} \left(-\frac{\pi }{3} x'\right)^{1/2}
H_{\frac{1}{6}}^{(1)}\left(- \frac{\sqrt{\beta_0}}{3}\,x'^3\right)
\ ,
\eeq
and then read off the power spectrum when the mode exits the horizon, $x'\to 0$. We have computed the factor $\gamma_s$ for several values of $\alpha_0$ and $\beta_0$. For example, for $\alpha_0=\beta_0=0.2$, one has $\gamma_s\simeq 0.717$, for $\alpha_0=\beta_0=0.5$, $\gamma_s\simeq 0.53$ and for $\alpha_0=\beta_0=1$, $\gamma_s=0.4$. It seems that in this case, $\gamma_s<1$ in most of the parameter space. In fact, solving and plotting the power spectrum for different values of $\alpha_0$ and $0.01<\beta_0<2$ confirms this conjecture. The larger the values of $\alpha_0$ and $\beta_0$, the more suppressed the power spectrum. However, this suppression is at most of order a few tenths. For $\alpha_0=\beta_0\sim 2$, the modulation factor $\gamma_s\approx 0.3$ (see Figs.~\ref{a0-b0-pos-1} and \ref{a0-b0-pos-2}). Noting that, in the case of positive quartic correction to  the dispersion relation, enhancement of the coefficient of the quartic term suppresses the power spectrum too~\cite{Ashoorioon:2011eg}, one is tempted to deduce that positive higher order corrections to the dispersion relation will in general reduce the amplitude of the power spectrum. 

\begin{figure}[t]
\includegraphics[width=\textwidth]{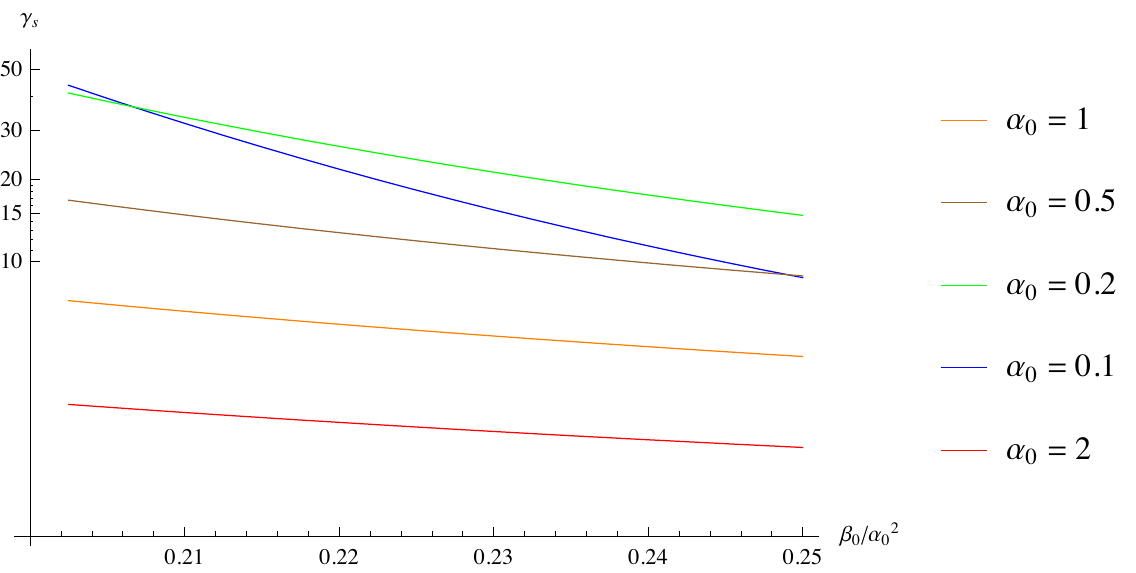}
\caption{Log plot of $\gamma_s$ as a function of $\beta_0/\alpha_0^2$ when $\alpha_0$ is big and negative ($0.1\leq|\alpha_0|\leq2$) and $\frac{\alpha_0^2}{5} \leq \beta_0 \leq \frac{\alpha_0^2}{4}$.
\label{Fig3}}
\end{figure}

\begin{figure}[b]
\includegraphics[width=\textwidth]{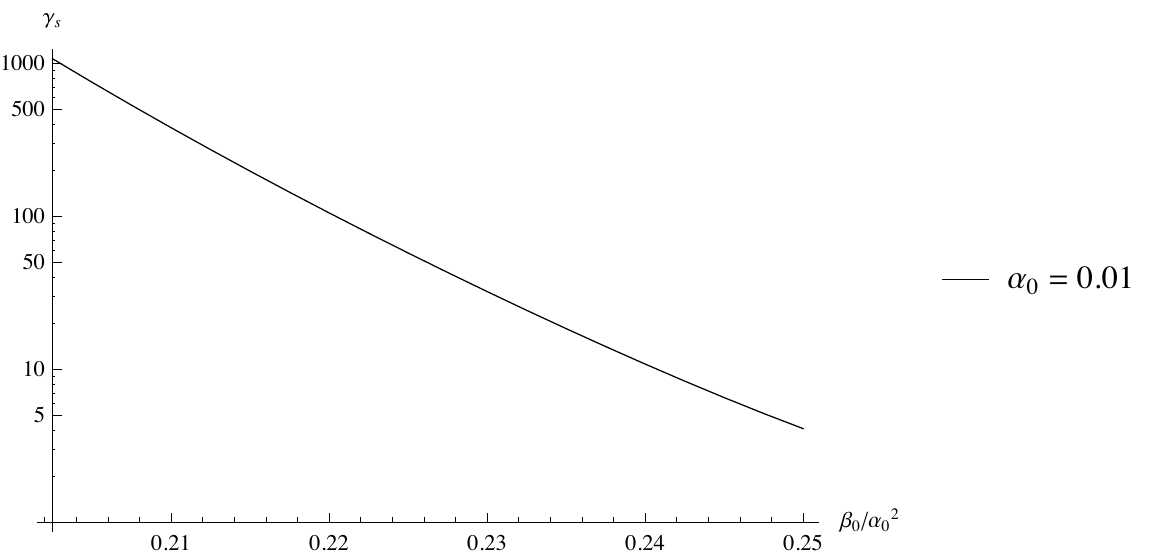}
\caption{Log plot of $\gamma_s$  as a function of $\beta_0/\alpha_0^2$ when $\alpha_0$ is small and negative ($|\alpha_0|=0.01$) and  $\frac{\alpha_0^2}{5} \leq \beta_0 \leq \frac{\alpha_0^2}{4}$.
\label{a0-neg-0.01}}
\end{figure}

On the other hand, if $\alpha_0<0$, there could be large modulation factors on the power spectrum. In~\cite{Ashoorioon:2017toq}, we focused on the range of parameters for $\alpha_0$ and $\beta_0$ such that 
\beq
z\equiv \frac{\beta_0}{\alpha_0^2}\geq \frac{1}{4}
\ .
\eeq
This requirement was coming from the fact that the dispersion relation should not become tachyonic for any physical momentum in flat spacetime. 
If we allow for $z<1/4$, the quantum modes will get amplified while they are still inside the horizon, which will increase the resulting power spectrum exponentially. However, such a large enhancement for the scalar power spectrum will probably come about at the price of enhancing the size of non-gaussianity, and may thus clash with experimental bounds on them. The issue of non-gaussianity with such modified dispersion relations is something that we plan to return to in the next work. As for the power spectrum itself, we examined this regime numerically. We have plotted in Figs. \ref{Fig3} and \ref{a0-neg-0.01} the factor $\gamma_s$, with $\beta_0$ in the range $\frac{\alpha_0^2}{5}<\beta_0<\frac{\alpha_0^2}{4}$, for several values of $\alpha_0$. As can be readily seen, when $\beta_0$ goes below $\alpha_0^2/4$, the power spectrum grows substantially. With the decrease of the parameter $\alpha_0$, the interval of time the mode spends under the influence of negative quartic contribution and possibly becomes tachyonic while still inside the horizon, increases and one sees a larger amount of enhancement for $\gamma_s$. In fact, for $\alpha_0=0.01$, and $\beta_0=\alpha_0^2/5$, $\gamma_s$ can reach values as large as 1000 (see Fig.~\ref{a0-neg-0.01}). One other point that may be worth highlighting is that for large values of $\beta_0$ for each $\alpha_0$, the maximum enhancement in this case seem to occur at $\alpha_0\sim 0.2$. As we will see, in the interval $1/4<z<1/3$, the same pattern repeats itself. 

\begin{figure}[t]
\includegraphics[width=\textwidth]{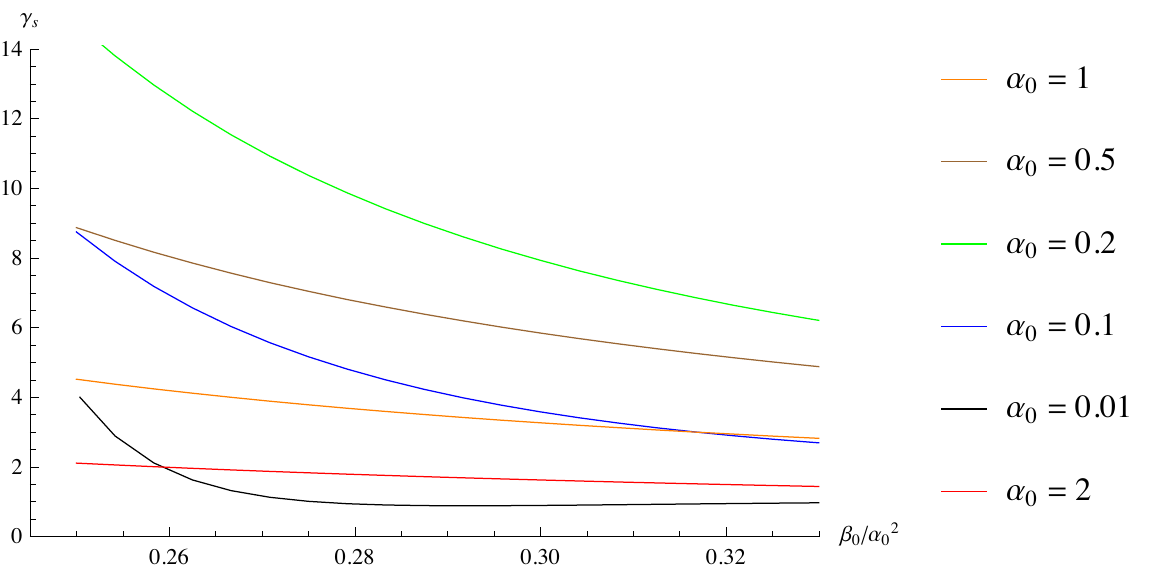}
\caption{$\gamma_s$ as a function of $\beta_0/\alpha_0^2$ when $\alpha_0<0$ and $\beta_0>0$ and $0.01 \leq |\alpha_0| \leq 2$ and $\frac{\alpha^2}{4} \leq \beta_0 \leq \frac{\alpha^2}{3}$.\label{Fig5}}
\end{figure}

\begin{figure}[t]
\includegraphics[width=\textwidth]{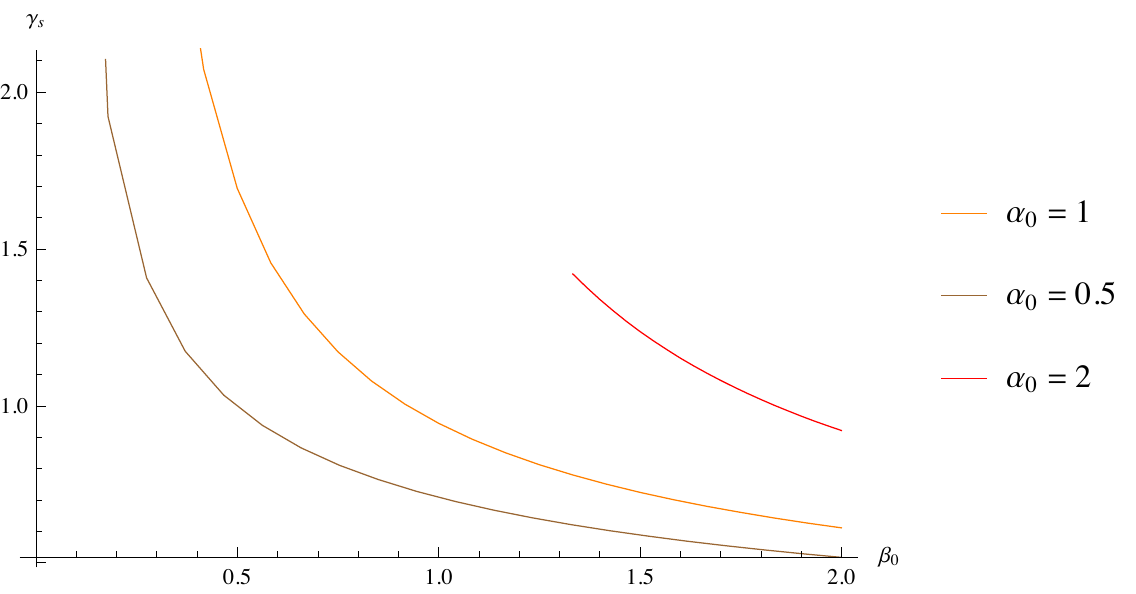}
\caption{$\gamma_s$ for big values of $\alpha_0$ (0.5, 1, 2) as a function of $\beta_0$ when $\alpha_0<0$ and $\beta_0>0$ and $\frac{\alpha^2}{3} \leq \beta_0 \leq 2$.\label{Fig6}}
\end{figure}

\begin{figure}[t]
\includegraphics[width=\textwidth]{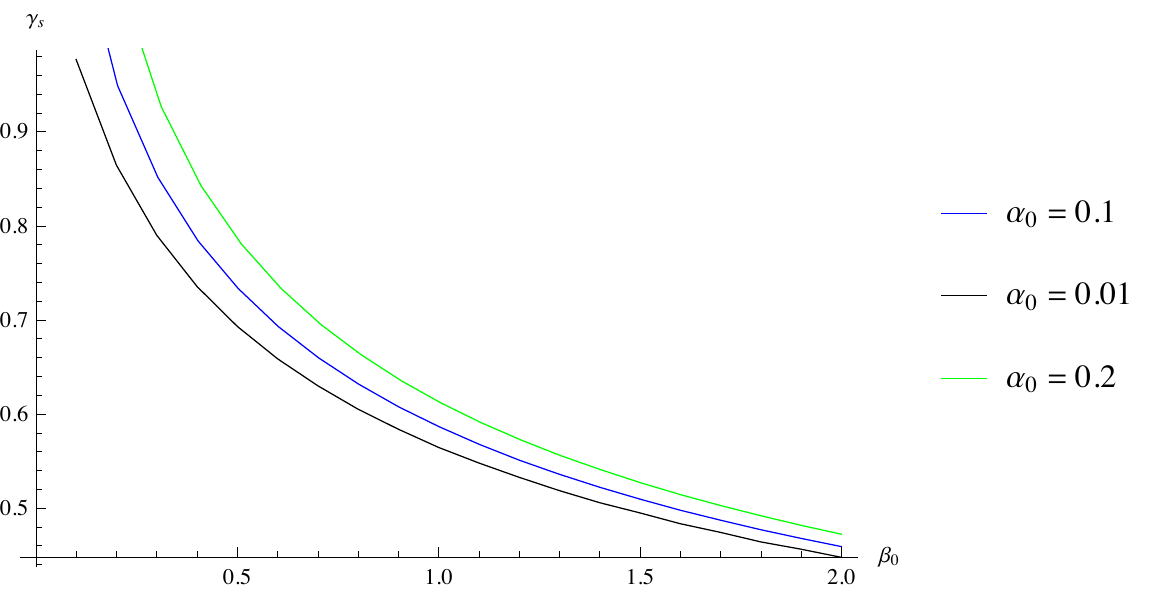}
\caption{$\gamma_s$ for small values of $\alpha_0$ (0.01, 0.1, 0.2)  as a function of $\beta_0$ when $\alpha_0<0$ and $\beta_0>0$ and $\frac{\alpha^2}{3} \leq \beta_0 \leq 2$.
\label{Fig7}}
\end{figure}

In~\cite{Ashoorioon:2017toq}, we had also assumed that the mode never becomes lighter than the Hubble parameter up until the last horizon crossing. That would mean that the equation $\omega^2=2H^2$ has only one solution. For 
\beq
z>\frac{1}{3}
\ ,
\eeq
this equation automatically has only one solution.  For values of $\az$ and $\bz$ such that
\beq
\frac{1}{4}\leq z \leq \frac{1}{3}\,,
\eeq
having only one solution imposes one of the following conditions
\beqa \label{az-range}
\az &\leq&  \frac{9 z-2-2 (1-3 z)^{3/2}}{54 z^2} \nonumber \\ 
\text{or} \quad \az &\geq& \frac{9 z-2+2 (1-3 z)^{3/2}}{54 z^2}\,.
\eeqa
The requirement of having one solution was to show that the gluing technique becomes ineffective in capturing the modification to the power spectrum from the sextic dispersion relation. In that case, since $\omega^2=2H^2$ had only one solution, calculating the power spectrum analytically required the gluing procedure only once. In this paper, we do not distinguish between multiple and single horizon crossings. What matters to us is the amplitude of the power spectrum when the wavelength of the mode is much larger than the Hubble length or, more quantitatively, when $x'\ll 1$. Therefore, when investigating the $z>1/4$ region, we allow for $z>1/3$ as well. Plots of $\gamma_s$ for different values of $\alpha_0$ and $\beta_0$ for $z>1/4$ are presented in Figs. \ref{Fig5}, \ref{Fig6} and \ref{Fig7}. In all cases, the factor $\gamma_s$ is bigger than one. As one moves away from $\az=0.2$ in both directions, the amount of enhancement in the power spectrum decreases.

Finally, in the case $z=1/4$, as noted in~\cite{Ashoorioon:2017toq}, for $\az\sim 0.2$ and $\bz={\az^2}/{4}$, one can achieve $\gamma_s\simeq 14.774$. 

\subsection{Tensor-to-scalar ratio and consistency relation}

Having modified spectra for both tensor and scalar fluctuations, it is expected  that the tensor-to-scalar ratio be modified as well. For general modified equations of motion for scalar and tensor perturbations, the tensor-to-scalar ratio is given by
\beq
r=16 \epsilon~ \frac{c_s}{c_\T} \frac{1}{\gamma_s}
\ .
\eeq
Therefore, a departure of any of the parameters, $c_s$, $c_\T$ and $\gamma_s$ from one, will bring about changes in the tensor-to-scalar ratio in comparison with the slow-roll counterpart. It is well known that the majority of kinetic term dominated inflationary models, known as K-inflation~\cite{ArmendarizPicon:1999rj}, would lower $r$. However, the possibility of enhancing $r$ by allowing for $c_s$ to be bigger than one has also been entertained in~\cite{Mukhanov:2005bu}. As mentioned before, whether superluminal propagation in EFTs is necessarily an indication of any pathology is still a topic of debate in the community~\cite{Goon:2016une}. 

Depending on the sign of the operator $-\frac{\bar M_3^2}{2}\, \delta K^{\mu}_{\ \nu}\delta  K^{\nu}_{\ \mu}$, which contributes to the speed of sound for gravity waves, one encounters two branches. In the {\it superluminal} branch, with $\bar M_3^2>0$, gravitons propagate faster than light during inflation, $c_\T>1$. In the {\it subluminal} branch, with $\bar M_3^2<0$, the speed of gravity waves is less than the speed of light. Were one to drop the superluminal branch because of superluminality and causality considerations, the effect of departure from the speed of light is always to enhance $r$. As $\bar M_3$ cannot be pushed beyond $M_{\rm Pl}$ in an EFT treatment, the gravity wave speed during inflation, is bounded from below by ${1}/{\sqrt{2}}$, and thus one at most would get an enhancement of $\sqrt{2}$ in $r$ in the subluminal branch. Otherwise, in the superluminal branch, $r$ can be significantly suppressed as $\bar M_3^2$ gets larger.

To summarize, in this section we showed that the effect of the EEFToI on scalar perturbations is to change their dispersion relation, not only by modifying the speed of sound but also by including higher order corrections. If the coefficients of the dispersion relation are all positive, the amplitude of the scalar power spectrum gets suppressed by a factor $\gamma_s\lesssim 1$. This would enhance $r$ by a factor of order one. However, if the quartic term of the dispersion relation is negative, the evolution of the modes involves an intermediate phase with negative group velocity, or tachyonic phase, which can in fact enhance the amplitude of scalar perturbations considerably with respect to the Bunch-Davies result, picking up a factor of $\gamma_s\gg 1$.  If one allows for the tachyonic evolution of the mode inside the horizon, $r$ can get suppressed by a large factor. On the other hand, if one only allows for having a negative group velocity without a tachyonic phase, the maximum one can achieve is $\gamma_s\approx 14.7$, which suppresses $r$ by a factor of $\sim 0.06$. This can be achieved by setting $\az\sim 0.2$ and $\bz=\az^2/4$. Latter was studied in~\cite{Ashoorioon:2017toq} as a dispersion relation that realises super-excited states with large occupation numbers. The mode equation has one turning point for such values of parameters, corresponding to a single horizon crossing. 

One can also easily verify that the consistency relation~\cite{Starobinsky:1985ww,Liddle:1992wi}
\beq\label{consistency}
n_\T=-\frac{r}{8}
\eeq
is modified to
\beq
n_\T=-\frac{c_\T \gamma_s}{c_s} \frac{r}{8}.
\eeq
This implies that for a given $r$, we will have a much redder tensor spectrum for $\gamma_s\gg 1$.

\section{Cut-off and non-gaussianity}
\label{NonG}

As explained in the introduction, the EEFToI can provide a controlled description of the perturbations around a quasi de Sitter background from low energy up to the UV regime where the dispersion relation becomes $\omega^2 \sim k^6$, provided the coefficients of the higher dimensional operators are tuned such that the theory remains always weakly coupled in this energy range. This means that the scattering of Goldstone bosons does not violate perturbative unitarity at any scale below the point where the dispersion relation becomes dominated by the sextic term. 

The strong coupling scale can be easily estimated in the $\pi$ language from the size of the coupling of the interactions described by higher dimensional operators \cite{Cheung:2007st}. In order to perform this estimate, let us first note that the canonically normalised Goldstone boson $\pi_c$ has a time kinetic term of the form
\beq
\int dt \,d^3 x \,\frac12\, \dot{\pi_c}^2\,, \nonumber
\eeq
and so the dimension of $\pi_c$, denoted as $[\pi_c]$, is
\beq
[\pi_c] = k^{3/2} \,\omega^{-1/2}\,.
\label{Pidim}
\eeq
The energy dimension of the coefficient $c$ of a generic operator $\mathcal{O}$,
\beq
\int dt \,d^3 x \,c\, \mathcal{O}\,, \nonumber
\eeq
can be inferred by imposing
\beq
\frac{c\,[\mathcal{O}]}{\omega k^3} = 1\,,
\label{rel}
\eeq
where $[\mathcal{O}] = k^s \omega^m [\pi_c]^p$ is the dimension of $\mathcal{O}$. Combining (\ref{Pidim}) and (\ref{rel}), we end up with
\beq
[c] = \omega^a k^b\qquad \text{where}\qquad a=1-m+\frac{p}{2}\qquad\text{and}\qquad b= 3-s-\frac{3p}{2}\,. \nonumber
\eeq
If we now use the dispersion relation $\omega^2 \sim k^{2n}$ we find
\beq
[c] = \omega^q\qquad\text{with}\qquad q = 1 -m + \frac{1}{n}\left(3-s\right)+ \frac{p}{2} \left(1 - \frac{3}{n}\right).
\eeq
Let us now consider, as illustrative examples, three different operators: $\mathcal{O}_1 = \dot\pi_c\left(\partial_i \pi_c\right)^2$, $\mathcal{O}_2 = \left(\partial_i \pi_c\right)^4$ and $\mathcal{O}_3 = \dot\pi_c^2 \left(\partial_i \pi_c\right)^2$ whose coefficients $c_i$ have energy dimensions $[c_i] = \omega^{q_i}$ with $i=1,2,3$. Different forms of the dispersion relation then give different energy scalings of these coefficients:
\begin{itemize}
\item $n=1$ $\qquad\Rightarrow\qquad$ $q_1 = -2 \qquad q_2 = -4 \qquad q_3 = -3$ 
\item $n=2$ $\qquad\Rightarrow\qquad$ $q_1 = -\frac14 \qquad q_2 = -\frac12 \qquad q_3 = -\frac54$
\item $n=3$ $\qquad\Rightarrow\qquad$ $q_1 = \frac13 \qquad \quad q_2 = \frac23 \qquad\,\,\,\, q_3 = -\frac23$ 
\end{itemize}
Interestingly, for $n<3$ we get $q_i<0$ $\forall\,i=1,2,3$, while $q_1$ and $q_2$ become positive for $n=3$. This implies that all three operators are \textit{irrelevant} when the dispersion relation is at most dominated by the quartic term. This behaviour can be shown to hold for all higher derivative operators. Hence the $\omega^2\sim k^2$ and $\omega^2\sim k^4$ theories have only UV strong coupling scales which we will denote respectively $\Lambda_2$ and $\Lambda_4$. On the other hand, for $\omega^2\sim k^6$, $\mathcal{O}_1$ and $\mathcal{O}_2$ become \textit{relevant}, while $\mathcal{O}_3$ remains \textit{irrelevant}. This implies that the $\omega^2\sim k^6$ theory features both a UV and an IR strong coupling scale which we will indicate respectively with $\Lambda_6^\UV$ and $\Lambda_6^\IR$. The scale $\Lambda_6^\UV$ is associated with operators like $\mathcal{O}_3$ whose coupling grows when the energy increases, whereas $\Lambda_6^\IR$ is related to operators like $\mathcal{O}_1$ and $\mathcal{O}_2$ whose strength grows when the energy decreases. In general, this behaviour is a signal of an inconsistent EFT description.

However, one can still have a sensible EFT for the fluctuations if $\Lambda_6^\IR$ is below the transition from the $\omega^2\sim k^4$ to the $\omega^2\sim k^6$ regime that happens at the scale $\Lambda_{\rm dis}^\46$, provided the $\omega^2\sim k^4$ theory is always weakly coupled in its regime of validity. Thus we need to tune the coefficients of the higher dimensional operators such that $\Lambda_6^\IR \ll \Lambda_{\rm dis}^\46 \ll \Lambda_4$. On top of this, the theory should not become strongly coupled at lower energies when the dispersion relation still takes the standard quadratic form. Hence we need also to require that $\Lambda_2$ is above the scale $\Lambda_{\rm dis}^\24$ which denotes the transition from the $\omega^2\sim k^2$ to the $\omega^2\sim k^4$ regime: $\Lambda_{\rm dis}^\24 \ll \Lambda_2$. 

The strong coupling scale $\Lambda_2$ can be estimated from the coefficient of the four-leg interaction term $\mathcal{O}_2$. Given that $q_2=-4$ in the deep IR where the dispersion relation is $\omega^2= c_s^2 k^2$, the coefficient $c_2$ scales as $c_2\sim 1/\Lambda_2^4$, where it can be shown that~\cite{Cheung:2007st}
\beq
\Lambda_2^4 = c_s^7 \Lambda^4 \qquad\text{with}\qquad \Lambda^4 = \frac{2 M_p^2 \dot{H}}{c_s^2\left(1-c_s^2\right)}\,.
\label{Lambda2}
\eeq
This strong coupling scale goes to infinity for $c_s\to 1$ but for $c_s\ll 1$, which is for example a regime interesting for large non-gaussianities, it can become very small. However, as noted in~\cite{Baumann:2011su}, similarly to particle physics, new physics is expected to appear before reaching the strong coupling regime. In turn, this new physics pushes the strong coupling region to higher energies. As explained above, in our case, new physics corresponds to a change in the dispersion relation from linear, $\omega^2=c_s^2 k^2$, to non-linear, $\omega^2 = k^4/\rho^2$, at the energy scale $\Lambda_{\rm dis}^\24 = c_s^2 \rho$. This happens before the $\omega^2 = c_s^2 k^2$ theory becomes strongly coupled if $c_s\,\rho^4 \ll \Lambda^4$. Moreover, in the region where the dispersion relation is quartic, the new strong coupling scale becomes $\Lambda_4^4 = \Lambda^4 (\Lambda/\rho)^{28}$~\cite{Baumann:2011su}. Hence the condition $c_s\,\rho^4 \ll \Lambda^4$ guarantees also that this strong coupling scale is indeed larger than the one in the case with a linear dispersion relation. 

As noted in~\cite{Cheung:2007st}, the strong coupling scale can also be quantified in terms of the size of non-gaussianities, since 
\beq\label{scoupled}
\frac{\mathcal{L}_3}{\mathcal{L}_2}\equiv f_\NL \zeta\,.
\eeq
Thus perturbation theory breaks down when this ratio becomes of order one. Given that observed value of $\zeta \sim 10^{-4}$, this happens if $f_\NL\sim 10^4$. However, notice that CMB observations set a stronger upper bound on equilateral non-gaussianities, of the order $f_\NL =-42\pm 75$~\cite{Ade:2013ydc}. The relation between non-gaussianities and strong couplings becomes even more manifest by noting that~(\ref{scoupled}) can also be rewritten as~\cite{Baumann:2011su}
\beq
\frac{\mathcal{L}_3}{\mathcal{L}_2}\sim \left(\frac{H}{\Lambda_2}\right)^2\,,
\label{scoupled2}
\eeq
which automatically results to $f_\NL \sim 10^4$, if horizon exit takes place in the strong coupling region, i.e. $H\sim \Lambda_2$. Given that $f_\NL \sim c_s^{-2}$, large non-gaussianities emerge when $c_s \ll 1$. From (\ref{Lambda2}) it is clear that small values of the speed of sound also imply a low strong coupling scale $\Lambda_2$. However if horizon exit occurs in a region where the dispersion relation is dominated by the quartic term, i.e. when $\Lambda_{\rm dis}^\24 \ll H\sim \Lambda_2 \ll \Lambda_4$, the theory should be weakly coupled. In fact, the ratio (\ref{scoupled}) takes an expression different from (\ref{scoupled2}) when evaluated in the $\omega^2 = k^4/\rho^2$ region
\beq
\frac{\mathcal{L}_3}{\mathcal{L}_2}\sim \left(\frac{H}{\Lambda_4}\right)^{3/4}\,.
\label{scoupled3}
\eeq
Given that $H\sim \Lambda_2 \ll \Lambda_4$, the theory is still weakly coupled even if $c_s\ll 1$. This implies also that $f_\NL$ should be smaller. In fact, the expression for $f_\NL$ in the quartic regime gets modified to $f_\NL \sim \left(\Lambda_{\rm dis}^\24/H\right) c_s^{-2}$ which is suppressed with respect to its expression in the quadratic regime since $\Lambda_{\rm dis}^\24 \ll H$. Hence we conclude that even a theory with small speed of sound and large non-gaussianities close to the edge of detectability can remain weakly coupled if the dispersion relation changes before hitting its strong coupling regime.

Notice also that, as explained in Sec.~\ref{Sps}, large non-gaussianities might arise also in the case where the coefficients of the higher dimensional operators are chosen such that $c_s\simeq 1$. In fact, in this case the transitions inside the horizon between regions with modified dispersion relations would give $f_\NL \sim \gamma_n / c_s^2$ with $\gamma_n$ which could potentially be large. This case might also be interesting to keep the EEFToI weakly coupled, since both $\Lambda_2$ and $\Lambda_4$ tend to infinity for $c_s\to 1$. However, one would still need to ensure that $\Lambda_6^\IR \ll \Lambda_{\rm dis}^\46$.

One can insist on pushing the strong coupling scale to even higher energies via the appearance of new physics again before reaching $\Lambda_4$. In this case, the new physics would correspond to a change in the dispersion relation from quartic to sextic at the scale $\Lambda_{\rm dis}^\46$ but we argued above that the $\omega^2\sim k^6$ regime features both an UV and IR strong coupling scale. Hence a sensible EFT description all the way up to the the sextic region requires to have also $\Lambda_6^\IR$ below $\Lambda_{\rm dis}^\46$. 

In addition to the scale where perturbative unitarity is lost, another fundamental scale for the EFT of the perturbations around a quasi de Sitter background is the scale $\Lambda_b$ where time diffeomorphisms are spontaneously broken by the background. Ref.~\cite{Baumann:2011su} showed that, for the standard slow-roll case with $c_s=1$, $\Lambda_b = \dot\phi^{1/2}\ll \Lambda_2$, while for $c_s \ll 1$, $\Lambda_2 \sim c_s \Lambda_b \ll \Lambda_b$. Moreover $\Lambda_b$ controls the size of the perturbations with respect to the background since
\beq
k^3 P_\zeta(k) =\frac14 \frac{H^2}{M_{\rm Pl}^2\,\epsilon\,c_s}\sim \left(\frac{H}{\Lambda_b}\right)^4\,,
\eeq
and so at energies around the symmetry breaking scale the background cannot be integrated out anymore to leave a theory for the fluctuations only. Hence we conclude that the cut-off of the EEFToI should be identified as the minimum between $\Lambda_6^\UV$ and $\Lambda_b$. As an illustrative example, a possible hierarchy of scales which would make the EEFToI valid for the entire energy range up to the $\omega^2\sim k^6$ region would be
\beq
\Lambda_{\rm dis}^\24 \ll \Lambda_2 \ll \Lambda_6^\IR \ll \Lambda_{\rm dis}^\46 \ll \Lambda_4 \ll \Lambda_6^\UV \ll \Lambda_b\,.
\eeq
We finally stress that cosmological observations are performed only at horizon crossing which occurs at energies of order $H$ where the dispersion relation might take the standard linear expression. Hence, strictly speaking, the only requirement to have a sensible EFT is $H\ll \Lambda_\cutoff$. However, we are demanding more since we want to provide a consistent theoretical framework for the picture described in~\cite{Ashoorioon:2017toq}. In fact, we want to be able to trust our EEFToI also in the high energy region which determines the behaviour of modes deep inside the horizon. We want these modes to be originally in the adiabatic vacuum of the $\omega^2\sim k^6$ theory, so that the standard $\omega^2\sim k^2$ theory features non-Bunch-Davis initial conditions due to the transitions between regions characterised by a different dispersion relation. In turn, these excited initial states can have very interesting implications for cosmological observables.

\section{Conclusions}
\label{Concl}

In this work we extended the formalism of the Effective Field Theory of Inflation, proposed earlier in~\cite{Cheung:2007st}, by including in the unitary gauge Lagrangian operators with mass dimensions three and four. There is a healthy combination of operators in this extended theory that allows one to avoid ghosts and imparts a sixth order correction to the dispersion relation for scalar perturbations. Modified dispersion relations are known to give rise to interesting effects on CMB observables due to non-Bunch-Davies initial conditions \cite{Ashoorioon:2017toq, Naskar:2017ekm, Choudhury:2017glj}. 

In this paper we provided a theoretical framework to motivate the emergence of modified dispersion relations. In particular, we discussed the conditions under which the Extended Effective Field Theory of Inflation is weakly coupled all the way from low energy to the UV region where the dispersion relation becomes dominated by the sextic term. In this situation, the evolution of the modes can be consistently followed from deep inside the horizon to super-horizon scales.

We also analysed the phenomenological implications of our EEFToI for the fluctuations around a quasi de Sitter background. We found that the tensor perturbations are untouched. On the other hand, various scenarios for the scalar perturbations can occur, depending on the parameters of the dispersion relation. If the coefficients of the quartic and the sextic dispersion relation are positive, the scalar power spectrum is suppressed. However, if the quartic coefficient is negative, even if for the case where the mode does not become tachyonic before crossing the only turning point, the scalar power spectrum can be enhanced. This in turn leads to a suppression of $r$ with respect to models with Lorentzian dispersion relation.

In this paper we focused on the implications of these models for two point function but in the future it would be interesting to also study the behaviour of the three point function, in particular, because the size of non-gaussianities control the strong coupling regime of our EFT. Hence, a careful analysis of non-gaussianities should provide a clear understanding on the possibility to tune the coefficients of the higher dimensional operators in order to have a sensible EFT description. This would correspond to non-gaussianities which are not larger than present observational constraints \cite{Ade:2013ydc}. Another issue we plan to return to in the future, is the investigation of the possibility to obtain further extensions of the EFToI with terms higher than $k^6$ in the dispersion relation.

\section*{Acknowledgement}

We are thankful to N.~Bartolo, C.P. Burgess, F.~Pedro, L.~Senatore, M.~M.~Sheikh-Jabbari, G. P. Vacca, M.~Zaldarriaga and in particular Paolo~Creminelli for many helpful discussions. A.A.~is thankful to the hospitality of Perimeter Institute where part of this work was completed. A.A.~and R.C.~are partly supported by INFN-FLAG.
G.G. and H.J.K. research is supported by the Discovery
Grant from Natural Science and Engineering Research Council of Canada (NSERC) and G.G.
is supported partly by Perimeter Institute (PI) as well. Research at PI is supported by the Government of
Canada through the Department of Innovation, Science and Economic Development Canada
and by the Province of Ontario through the Ministry of Research, Innovation and Science.

\bibliographystyle{JHEP}
\bibliography{bibtex}

\end{document}